\documentclass[twocolumn, 
               astrosymbooll,
               paper]{aastex631}
\usepackage{amsmath}

\newcommand{\kp}{$\kappa$}
\newcommand{\g}{$\gamma$}
\newcommand{\s}{$s$}
\newcommand{\allmaps}{$12,342$}
\newcommand{\ger}{GERLUMPH}
\newcommand{\LSR}{latent space representation}
\newcommand{\set}{(\kp, \g, \s)}
\newcommand{\AD}{AD}
\newcommand{\BCE}{BCE}
\newcommand{\KL}{KL}
\newcommand{\RE}{$R_E$}
\newcommand{\testLC}{\ensuremath{\mathrm{test}_{LC}}}

\newcommand{\Gmap}{\ensuremath{\mathrm{\it Gmap}}}
\newcommand{\ADmap}{\ensuremath{\mathrm{\it ADmap}}}
\newcommand{\bestAD}{{\it AD50-BCE-KL}}
\newcommand{\FIDmodel}{\texttt{inceptionV3}}
\newcommand{\ugrizy}{{\it ugrizy}}

\usepackage{soul}

\begin{document}

\title{Autoencoder Reconstruction of Cosmological Microlensing Magnification Maps}

\author[0000-0002-1910-7065]{Somayeh Khakpash}
\affiliation{Rutgers University--New Brunswick, Department of Physics \& Astronomy, 136 Frelinghuysen Rd, Piscataway, NJ 08854, USA}
\affiliation{LSST-DA Catalyst Fellow}

\author[0000-0003-1953-8727]{Federica~B.~Bianco}
\affiliation{University of Delaware
Department of Physics and Astronomy
217 Sharp Lab
Newark, DE 19716 USA}
\affiliation{University of Delaware
Joseph R. Biden, Jr. School of Public Policy and Administration, 
184 Academy St, Newark, DE 19716 USA}
\affiliation{University of Delaware
Data Science Institute}
\affiliation{Vera C. Rubin Observatory, Tucson, AZ 85719, USA}

\author[0000-0001-8554-7248]{Georgios Vernardos}
\affiliation{Department of Physics and Astronomy, Lehman College of the City University of New York, Bronx, NY, 10468, USA}
\affiliation{Department of Astrophysics, American Museum of Natural History, Central Park West and 79th Street, NY, 10024, USA}

\author[0000-0002-9276-3261]{Gregory Dobler}
\affiliation{University of Delaware
Department of Physics and Astronomy
217 Sharp Lab
Newark, DE 19716 USA}
\affiliation{University of Delaware
Joseph R. Biden, Jr. School of Public Policy and Administration, 
184 Academy St, Newark, DE 19716 USA}
\affiliation{University of Delaware
Data Science Institute}

\author[0000-0001-6812-2467]{Charles Keeton}
\affiliation{Rutgers University--New Brunswick, Department of Physics \& Astronomy, 136 Frelinghuysen Rd, Piscataway, NJ 08854, USA}

\begin{abstract}

Enhanced modeling of microlensing variations in light curves of strongly lensed quasars improves measurements of cosmological time delays, the Hubble Constant, and quasar structure. Traditional methods for modeling extra-galactic microlensing rely on computationally expensive magnification map generation. With large datasets expected from wide-field surveys like the Vera C. Rubin Legacy Survey of Space and Time, including thousands of lensed quasars and hundreds of multiply imaged supernovae, faster approaches become essential. We introduce a deep-learning model that is trained on pre-computed magnification maps covering the parameter space on a grid of \kp, \g, and $s$. Our autoencoder creates a low-dimensional latent space representation of these maps, enabling efficient map generation.  Quantifying the performance of magnification map generation from a low dimensional space is an essential step in the roadmap to develop neural network-based models that can replace traditional feed-forward simulation at much lower computational costs. We develop metrics to study various aspects of the autoencoder generated maps and show that the reconstruction is reliable. Even though we observe a mild loss of resolution in the generated maps, we find this effect to be smaller than the smoothing effect of convolving the original map with a source of a plausible size for its accretion disk in the red end of the optical spectrum and larger wavelengths and particularly one suitable for studying the Broad-Line Region of quasars. 
Used to generate large samples of on-demand magnification maps, our model can enable fast modeling of microlensing variability in lensed quasars and supernovae.

\end{abstract}


\section{Introduction} \label{sec:intro}

In a strong lensing system, the gravity from massive galaxies between us and a distant astronomical source can create multiple resolved images of that object. Additionally, the magnification of each image of the source will be modified by compact objects like stars in the lens galaxy at the location of the image, causing microlensing variability in the light curves of the multiple images of the source \citep{paczynski1986gravitational, kayser1986astrophysical}. 
Studying the microlensing behavior in multiply-lensed distant objects like quasars and supernovae is important from various perspectives. From a cosmological point of view, measures of time delays of multiple images, from which one can derive the Hubble constant, would be enhanced by perfect removal of the microlensing variability in the light curves \citep{dobler2006microlensing, tewes2013cosmograil, aghamousa2015fast, jackson2015hubble, 
dobler2015strong, bonvin2016cosmograil, tsvetkova2016simple, chen2018constraining, tie2018microlensing, hu2020modeling, millon2020cosmograil, leon2023data, meyer2023td, birrer2024time}. On the other hand, modeling the microlensing variability would help to study the size and brightness profile of a quasar \citep{blackburne2011sizes, vernardos2019quasar,cornachione2020microlensing, 
fian2021revealing, fian2021microlensing, chan2021measuring,
paic2022constraining,
fian2023diffuse,
savic2024probing,vernardos2024microlensing}, the structure of the expanding photosphere of a supernova \citep{dobler2006microlensing,suyu2024strong}, and also the mass distribution of stars and compact objects within the lens galaxies \citep{wyithe2000limits, rusin2001constraints, wyithe2001determining, winn2003investigation, dobler2007microlensing,  schmidt2010quasar, schechter2014stellar, oguri2014stellar, jimenez2019initial, esteban2020impact,  tuntsov2024free}. The microlensing effect in strongly lensed objects also depends on the ratio of smoothly distributed dark matter to total mass, providing an opportunity to study the distribution of dark matter across a galaxy and compare it to the results drawn from dynamical studies of galaxies and simulations of galaxy formation \citep{schmidt1998limits, schechter2004dark, congdon2007microlensing, schmidt2010quasar, bate2011microlensing, jimenez2015dark, fedorova2016gravitational, mediavilla2017limits, esteban2020impact}.

The Vera C. Rubin Observatory Legacy Survey of Space and Time \citep[LSST,][]{ivezic2019lsst} will observe the whole southern sky over 10 years, taking two to three observations every two to three days in six bands (\ugrizy). In the era of the Rubin LSST, the number of discovered multiply-imaged quasars and supernovae will increase by more than a factor of 10 \citep{OguriMarshall2010, Lemon2024grav, 2024MNRAS.531.3509A} providing several thousands of events, a golden opportunity to make great advancements in our scientific understanding of these systems.
To model the microlensing variability on a larger number of objects, fast, accurate, and efficient methods are required. 

\begin{figure}
    \centering
    \includegraphics[width=0.9\columnwidth]{ 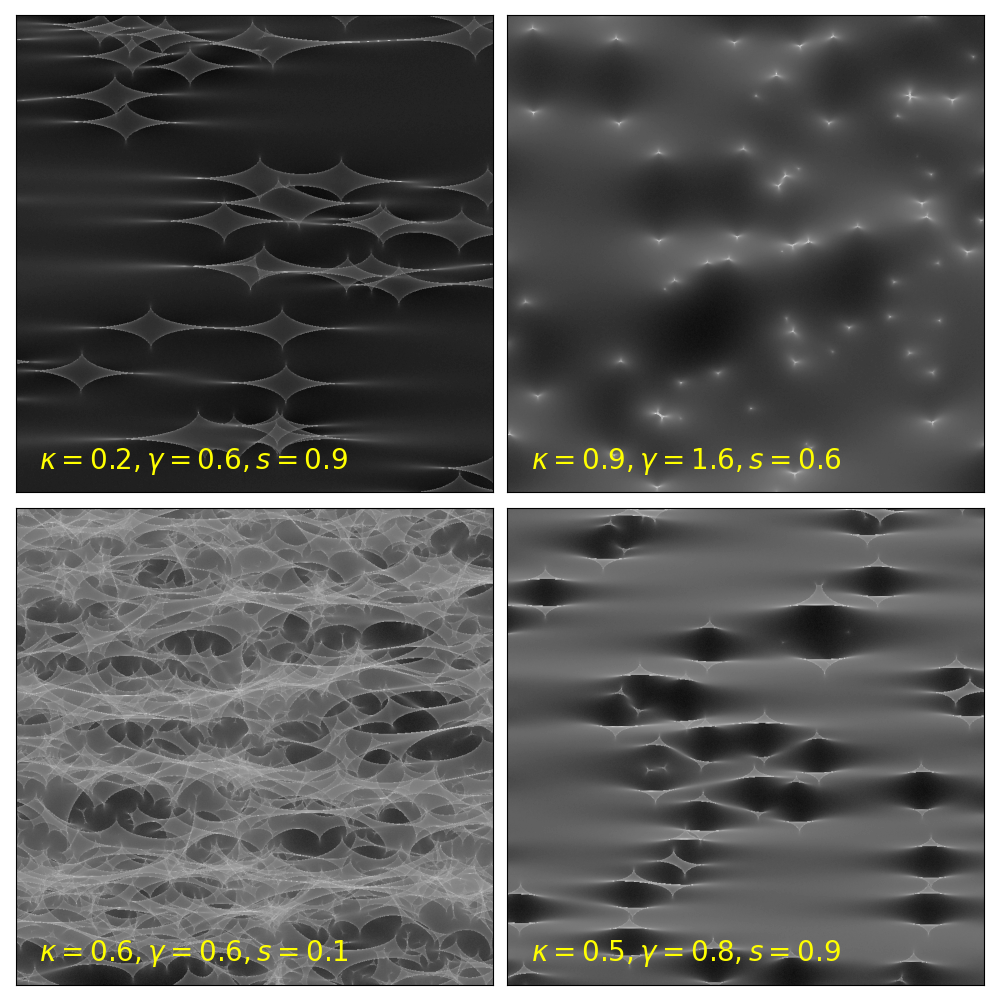}
    \caption{Four instances of \ger\ maps with different values of \kp, \g, and \s. This figure is discussed in \autoref{sec:ger_maps}.}
    \label{fig:map_examples}
\end{figure}

The effect of the mass distribution of the part of the lens galaxy along the line of sight is calculated in the form of a magnification map (see \autoref{fig:map_examples}). The parameters of the lens galaxy include \kp\ (total mass density), $\gamma$ (shear due to external field, meaning the rest of the galaxy), and $s$ (smooth dark matter fraction), which, along with the stars' positions, determine what the magnification map looks like (see also \autoref{sec:ger_maps}). As seen in \autoref{fig:map_examples}, the number and shape of the caustics (regions of high magnification) and de-magnification regions can significantly change with different values of \set.

The trajectory of the source along the magnification map determines how its brightness will change due to the microlensing effects. The microlensing variability, along with the intrinsic variability of the lensed quasar, can be observed in the light curve of the quasar.
When modeling the microlensing variability in light curves of multiply-imaged quasars or supernovae, one fits for parameters of the lens galaxy describing the distribution of compact objects and smooth dark matter in the galaxy, parameters of the source like the size and brightness profile of the quasar/supernova, and the trajectory of the source in the lens plane \citep{vernardos2024microlensing}. In an ideal scenario, to infer the parameters of the lens galaxy and the source from the light curves, all possible trajectories in the parameter space of all possible magnification maps of the lens galaxy should be probed to find the best fit to the observed light curves. However, modeling and interpreting the microlensing variability in the light curves of a multiply-imaged quasar/supernova is very challenging because different sets of parameters of the lens-galaxy mass distribution can create similar features in the light curves. Note that low-cadence observations and short-time baseline of the light curves can significantly worsen the constraints on the model. 

Generating high-resolution magnification maps covering a wide enough area on the lens galaxy is usually done by inverse ray-shooting, which requires high processing time and memory \citep{kayser1986astrophysical, vernardos2014adventures}. With new advancements in hardware technology, the generation of magnification maps has become achievable, and therefore, datasets of pre-computed maps have become available. \citep{vernardos2014gerlumph, shalyapin2021fast}. 



One of the most recent advances in the field is the generation of the \ger \footnote{\url{https://gerlumph.swin.edu.au}} magnification maps \citep{vernardos2014gerlumph}. This is a collection of pre-computed maps that are stored at an online server and can be retrieved upon submitting queries for each set of \set. 
\ger\ has significantly enhanced the process of modeling microlensing variability, however, accessing to large samples of maps and studying their large-scale features of them remains challenging.   \citet{shalyapin2021fast} and \citet{zheng2022improved} also introduced approaches to generating magnification maps faster and more efficiently. \citet{shalyapin2021fast} have implemented the code on an online server\footnote{\url{https://microlensing.overfitting.es}} where maps can be generated on demand for arbitrary sets of parameters. They provide maps with a maximum image size of 2,000$\times$2,000 pixels and physical size ranging  from $L = 5~R_E$ to $L=100 ~R_E$, where $1\; R_E$ is the Einstein radius of a 1 $M_{\odot}$ star. \citet{zheng2022improved} introduced a method that significantly improves the time needed for generating 10,000$\times$10,000 maps. Finally, to make these large computations feasible when generating the maps, stars are assumed to be static. In reality, we should have time-dependent maps due to the relative movement of the stars and compact objects within the lens galaxy to better study the effect of that on the microlensing variability \citep{vernardos2024microlensing}.


\begin{figure}
    \centering
    \includegraphics[width=0.9\columnwidth]{ 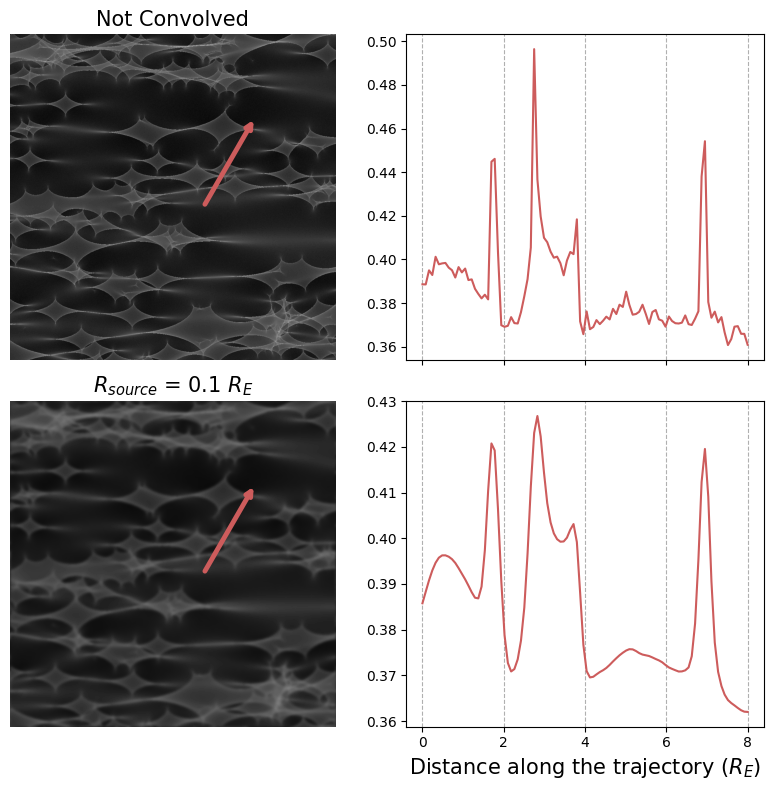}
    \caption{{\it Left Column}: A magnification map with parameters \kp$=0.3$, \g$=0.5$, and \s$=0.5$. The red line shows the trajectory of the source through the maps. The map in the first row is not convolved, and the map in the second row is convolved with a source size of $0.1 \; R_E$. {\it Right column:} The light curves show the microlensing variability of a source traveling along the trajectories indicated in the maps. The $x$ axis of the light curves are distances along the trajectories in units of $R_E$. Note that the  map's physical scale is 25$R_E$ $\times$ 25$R_E$.}
    \label{fig:map_lc_examples}
\end{figure}

Our goal is to create  an accessible framework to easily and quickly generate ensembles of magnification maps without having to store the maps locally to use in modeling and simulation. As a first step in this process, we introduce an autoencoder (\AD) model \citep{bank2020autoencoders}, which is a deep artificial neural network trained to reproduce its own input. We trained our model on a lower resolution version of \allmaps\ \ger\ Data Release 1 maps covering the parameter space of \set\ on a grid \citep{vernardos2014gerlumph}. This model learns to generate a lower dimensional representation of the maps of size $50 \times 50$ pixels from maps of  $1,000\times1,000$ pixels (hereafter \Gmap s) and to reconstruct them from the lower dimensional space. In this paper, we show that the  maps generated by the \AD\ (hereafter \ADmap s) are a suitable replacement for the \ger\ maps for observations in the red end of the optical spectrum ($r$ filter and redder) and larger wavelengths and in particular for studying Broad-Line Region (BLR) of the quasar. Importantly, in order to demonstrate that the \ADmap s can replace the original maps, we have developed a set of novel metrics that can serve as foundations for evaluating future machine learning models used on magnification maps.


In \autoref{sec:ger_maps}, we introduce the magnification maps used as the training set. We then introduce and discuss the \AD\ architecture and training in \autoref{sec:AD} and \autoref{sec:training}, respectively. In \autoref{sec:results}, we show examples of the reconstructed maps using the \AD\ and the light curves generated from them. In \autoref{sec:metrics},  we introduce the metrics we have developed to evaluate the fidelity of the reconstructed maps. Finally, in \autoref{sec:concl}, we summarize and conclude.

\begin{figure*}[ht!]
    \centering
    
    \includegraphics[width=0.95\textwidth]{ 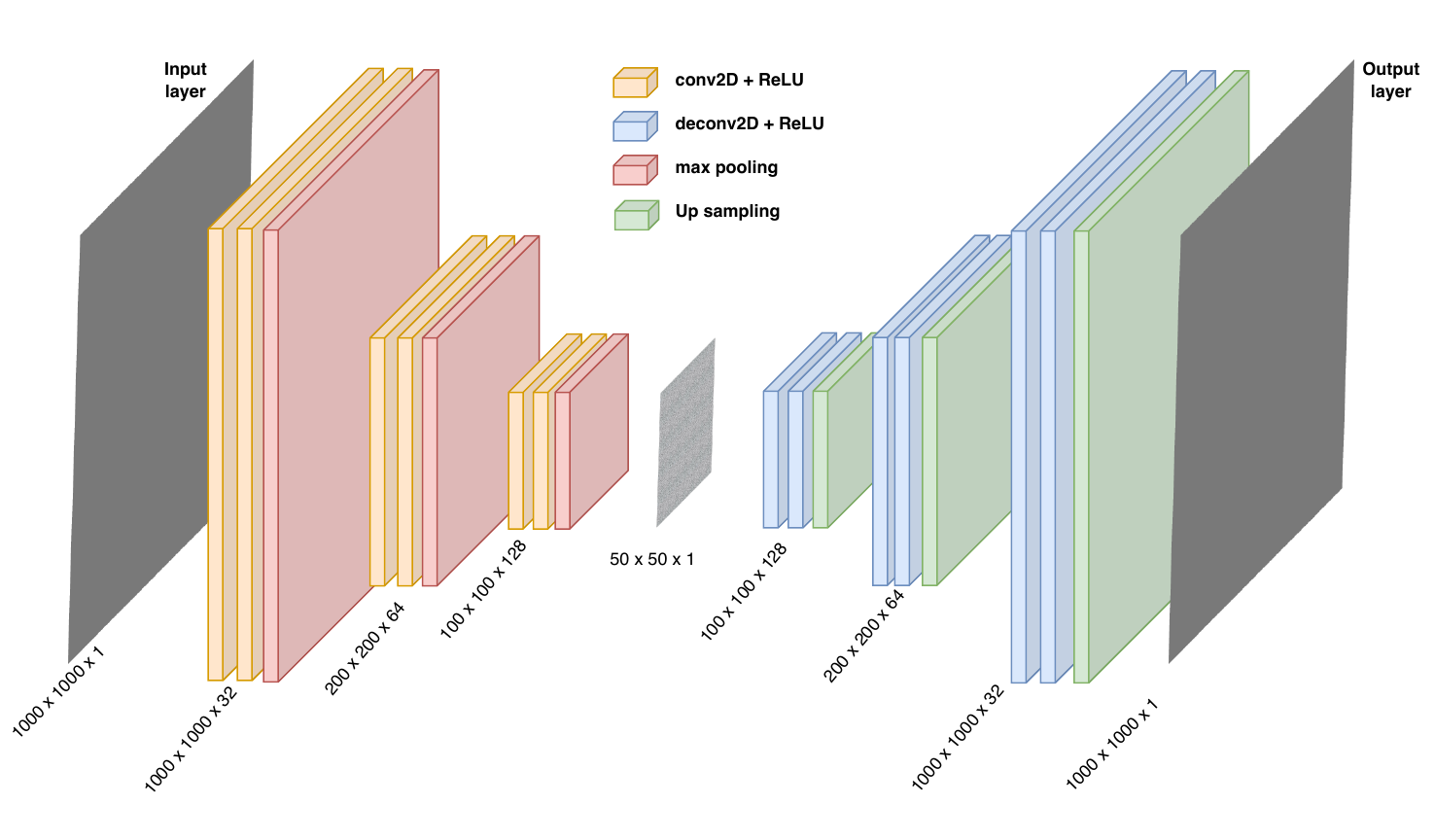}
    \caption{The structure of our \AD\ designed to reproduce \Gmap s. This architecture corresponds to the models labeled as {\it AD50} in \autoref{tab:trials}.} \label{fig:AD_structure}
\end{figure*}

\section{Map Reconstruction} \label{sec:maps_recon}

\subsection{\ger\ Maps}\label{sec:ger_maps}
GPU-Enabled High-Resolution cosmological Microlensing parameter survey (\ger) is a collection of pre-computed magnification maps that serve as training data for our model. Each map represents the overall gravitational effects of a distribution of compact objects within a small patch on the lens galaxy as seen on the source plane \citep{vernardos2014gerlumph}. There are three main parameters that describe the mass distribution in a galaxy and, therefore, determine what the magnification maps will look like. These parameters are \kp\ which is the ``convergence'' and describes the total mass. The parameter \g\ is the shear due to the external mass and it describes the overall gravitational effect of external mass on the magnification maps. The parameter \s\ is the smooth dark matter fraction and determines how much of the total mass is in the form of smooth dark matter as opposed to compact objects. The positions of the stars are chosen randomly, and they all have the same mass of $M_s = 1\; M_{\odot}$. It has been shown that choosing a varying set of masses for the microlenses has little effect on the microlensing variability \citep{wambsganss1992probability, lewis1995statistics, wyithe2002constraints, congdon2007microlensing}. It is important to note that with the same \kp, \g, and \s, if the stars' positions change, large enough maps will have similar statistical properties \citep{vernardos2013new}, but they will not be exactly the same.

\autoref{fig:map_examples} shows four examples of \ger\ magnification maps for different sets of \set. These maps are $10,000 \times 10,000$ pixels and cover an area of $25 \times 25 \; R_E$. Source trajectories across these maps give rise to hypothetical light curves. The first row of \autoref{fig:map_lc_examples} shows an example of a light curve (right panel) generated for a trajectory (red line) along a magnification map (left panel) assuming a point source. However, when observed, due to its finite size, an object cutting a trajectory across the caustic map will suffer magnification corresponding to a convolution of the caustic map with a kernel that depends on the source light distribution. We use a Gaussian kernel to describe the brightness profile of a simulated source and convolve that with the maps. The source profile shape does not significantly change the microlensing variability, and a Gaussian kernel is a good approximation of the source brightness profile \citep{Mortonson2005Size, vernardos2024microlensing}. 
We quote the standard deviation of the Gaussian kernel as the size $R_{source}$ and consider two values $R_{source} = 0.1\ R_E,\; 0.5\; R_E$. These values of source sizes are reasonable in the red end of the optical spectrum and lower frequencies \citep{pooley2007x}. When studying BLR of the quasars, these sizes will be even larger. In \autoref{fig:map_lc_examples}, the second row shows the same light curves for the same trajectory on the same map when the source size is $0.1\; R_E$. 

\begin{table*}
\begin{center}

\caption{Architecture and hyperparameter differences in selected \AD\ models.}\label{tab:trials}
\centering
\begin{tabular}{ |l|c|c|c|c|c|c| } 
 \hline
 
 & {\it AD50-TW }& {\it AD25-BCE }& {\it AD50-BCE}& {\it AD50-BCE }& {\it AD50-BCE-KL }& {\it AD50-BCE-KL}\\ 
 & & & {\it large batch} & & {\it short} &  \\ 
  \hline
 Latent Space Representation (LSR) Size & $50\times50$ & $25\times25$ & $50\times50$ & $50\times50$ & $50\times50$ & $50\times50$\\
 Training Set Size & 3828 & 12300 & 12300 & 12300 & 12300 & 12300  \\
 Epochs & 150 & 50 & 300& 300 & 50 & 300\\
 Learning Rate (LR) & $10^{-5}$ &  $10^{-5}$ & $10^{-5}$ & $10^{-5}$ & $10^{-5}$ & $10^{-4}$\\
 Loss Function & TW & BCE & BCE & BCE & BCE-KL & BCE-KL\\
 Batch Size & 8 & 8 & 16 & 8 & 8 & 8\\
 
 \hline
\end{tabular}
\end{center}
\end{table*}

\subsection{The Autoencoder (AD)} \label{sec:AD}

An \AD\ is a deep-learning-based model that reduces the dimensions of the input images while preserving the information in a way that would enable it to then reconstruct the input images from the reduced representation \citep{bank2020autoencoders}. The target of the encoder is either identical to or an enhanced version of the input (for example, to increase image resolution or denoise data). In our case, we pass \ger\ maps as input \emph{and} output. The smallest dimension layer of the AD (called the \LSR) forms a small meaningful parameter space that contains the most important information in the input, enabling the reconstruction. 

The \AD\ has two main parts (see \autoref{fig:AD_structure}): the encoder and the decoder. The encoder is a neural network, convolutional in our case. It runs learnable convolution kernels of different sizes over the input image in several steps to extract relevant features and decrease the dimension of the images gradually using layers called MaxPooling2D in the tensorflow package of Python. The MaxPooling2D layers replace the maximum pixel value of all pixels in a small window of arbitrary size with the original pixels and have the effect of reducing noise in the model while shrinking the input.  The \LSR\ is the output of the encoder and the encoded information that will be passed on to the decoder as input. The decoder usually has the same structure as the encoder but in reverse, aiming at increasing the dimensionality. So, in our model, it performs transpose convolution operations within a set of layers and increases the dimensions of the input through UpSampling2D layers (layers in the tensorflow package that work in reverse of the MaxPooling2D to increase dimensions.). 

The structure of our baseline \AD\ includes three blocks of double 2D convolutional layers ---kernel size = $(3, 3)$--- of the same number of filters ( 128, 64, and 32 for the three blocks, respectively) followed by a MaxPooling2D layer on the encoder and three blocks of double transpose 2D convolutional layers of the same size followed by an UpSampling2D layer on the decoder. A diagram of the structure is shown in \autoref{fig:AD_structure}.

We test the performance of two versions of the \AD\  with different latent space sizes: $50\times50$ and $25\times25$ pixels, and different sets of training hyperparameters for the $50\times50$ version. In total we test six versions of the \AD\ as described in \autoref{tab:trials}. The most effective \AD\ architecture is named ``AD50-BCE-KL''.  We will compare their results in \autoref{sec:results}.

\subsection{Training} \label{sec:training}

We have trained and tested a set of \AD s on a set of \allmaps\ \ger\ maps, and we evaluate their performance from various perspectives. The \ger\ maps are large (0.4 GB each), and working on the original version requires too much memory. We decrease their size to $1,000 \times 1,000$ pixels by choosing every 10th pixel in the original maps. Hereafter, we refer to the original $10,000 \times 10,000$ pixel maps as \ger\ maps and the lower resolution $1,000 \times 1,000$ pixel maps as \Gmap s. We will demonstrate in \autoref{sec:performance} that while we train on reduced resolution maps, our work extends naturally to the full resolution versions.

\begin{figure*}[ht!]
    \centering
    \includegraphics[width=0.9\textwidth]{ 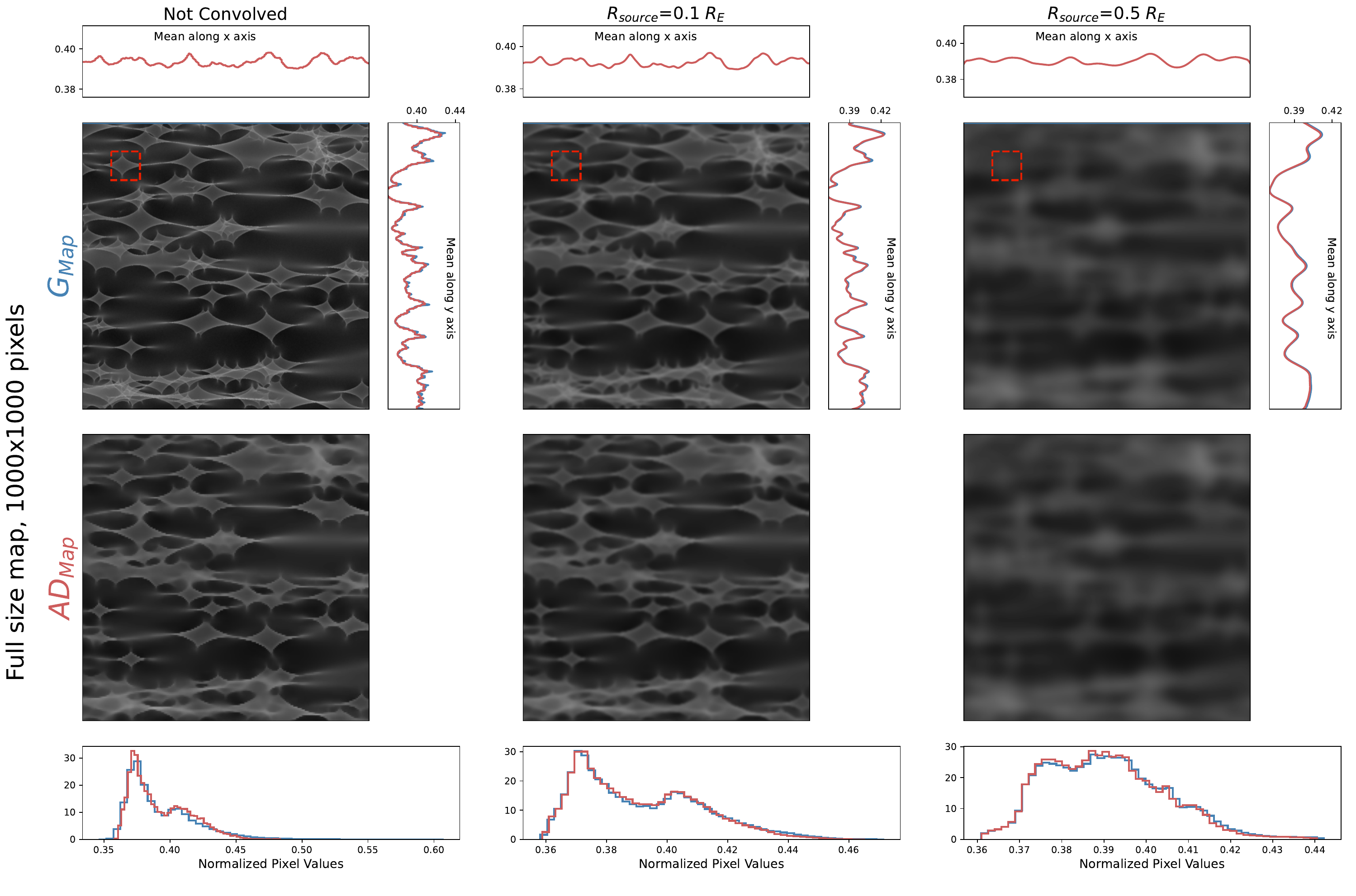}
    \caption{ An example \Gmap\ with parameters $\kappa = 0.3, \gamma = 0.5, {s} = 0.5$ is reconstructed using the \AD. The second and third rows show the \Gmap s and the \ADmap s, respectively. The panels above (first row) and to the right of each \Gmap\ show the mean of pixel values along the columns and rows respectively. The fourth rows are histograms of pixel values of the \Gmap s  (blue) and the \ADmap s (red). Each column corresponds to a convolution status of the maps and their reconstructions so that the first column represents the unconvolved maps, and the other columns correspond to the convolution with a source size of $ R_{source} = $ 0.1 and 0.5 \RE.  }
    \label{fig:compare_results}
\end{figure*}

\ger\ map pixel values are originally in units of ray counts. For each map, an average magnification and ray count are also given as metadata. We use those to convert ray counts to magnification as in
\begin{equation}
\label{eq:raycount}
    {\mu}_{ij} = N_{ij}\frac{\langle\mu\rangle}{\langle N\rangle},
\end{equation}
where $N_{ij}$ and ${\mu}_{ij}$ are the number of ray counts and magnification in each pixel, respectively, and $\langle N \rangle$ and $\langle \mu \rangle$ are average ray count and magnification for each image. We then calculate ${\mu^\prime}_{ij}=\log({\mu}_{ij}+0.004)$ (where $0.004$ is added to avoid having magnifications of zeros in the logarithm and this value is chosen as the minimum magnification value larger than zero in the maps), and normalize the maps to be between 0 and 1 by a min-max normalization  Values for all maps are normalized the same way based on the minimum and maximum of ${\mu^\prime}_{ij}$ in all of the maps ($\mu_{ij\mathrm{\_norm}}' = \frac{{\mu^\prime}_{ij} +3}{9}$).


The job of the \AD\ is to reconstruct the input \Gmap. The reconstructions are called \ADmap s. By showing that the \AD\ reconstruction performance is acceptable for all maps in the parameter space of \set\ we implicitly demonstrate that the \LSR\ contains the most important and relevant information in the maps and what was discarded in the dimensionality reduction was noise.

The networks are trained with different loss functions, for different numbers of epochs, with different values of learning rates and batch sizes. 

\begin{figure*}[ht!]
    \centering
    \includegraphics[width=0.9\textwidth]{ 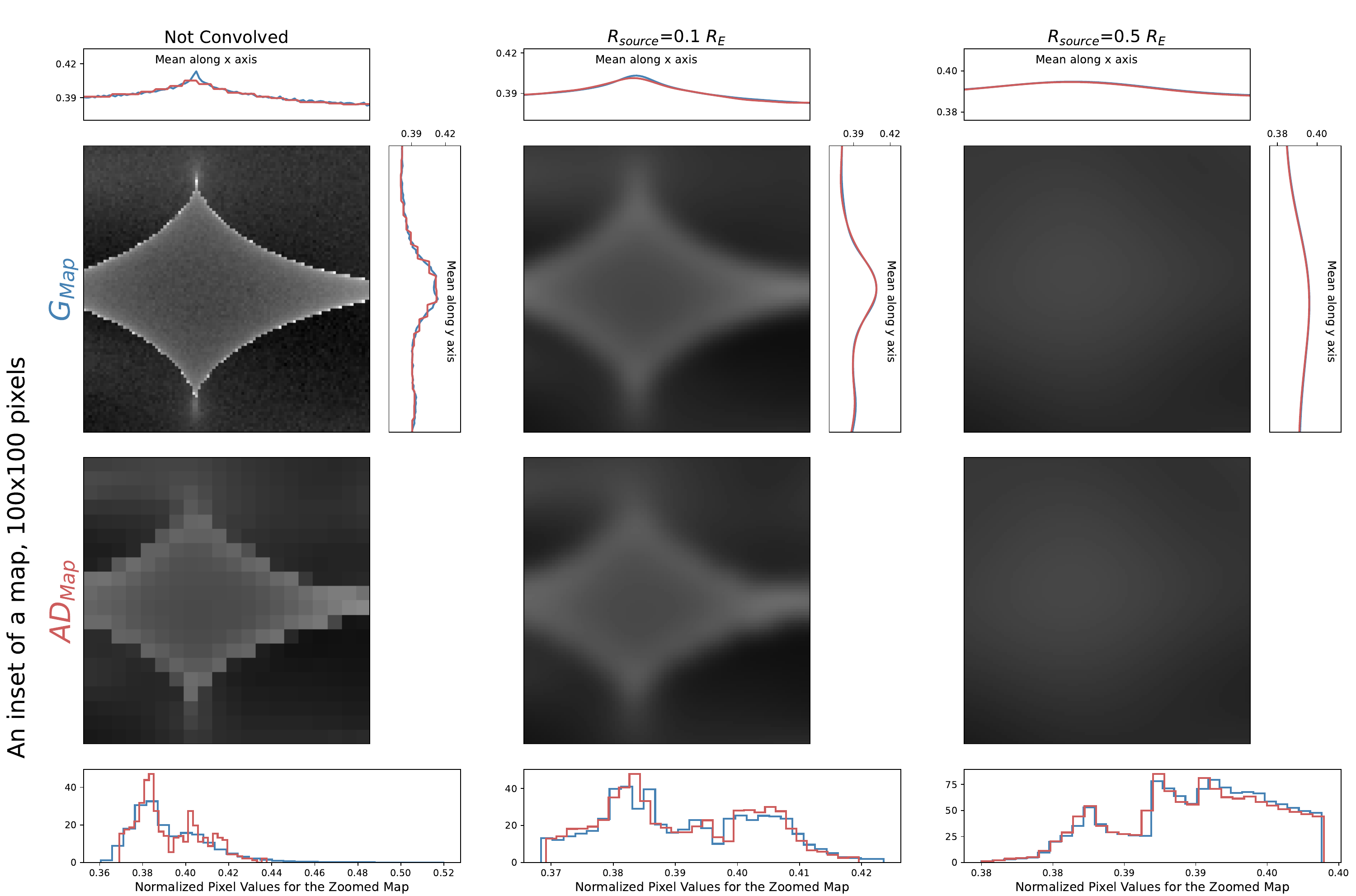}
    \caption{Same as \autoref{fig:compare_results}, but for a $100\times100$ pixel portion of the example map (indicated in the first row of \autoref{fig:compare_results} by the red dashed square).}
    \label{fig:compare_results_zoom}
\end{figure*}

\subsection{Loss functions}\label{sec:loss}
The loss function is a key component of the architecture and the loss functions we tested in our work are described below.

For reconstruction using \AD s, a mean squared error is commonly used. However, in cases with black-and-white high-contrast images where pixel values are either very low or very high, a binary cross entropy (\BCE), typically used for binary classification, can be useful\footnote{Note that, while it is designed for binary outputs, the \BCE\ returns a probability for each output, thus its output is in fact continuous. }. The \BCE\ is defined in
\begin{equation}\label{eq:bce}
BCE\left(x, y\right) = \frac{-1}{N}\sum_{i,j}^{} \left[ x_{ij}\log \left(y_{ij} \right)+ \left(1-x_{ij} \right)\log \left(1-y_{ij}\right) \right]
\end{equation}
where $x_{ij}$ and $y_{ij}$ are the input and output pixel values respectively, and $N$ is the total number of pixels.

We found that adding a Kullback-Leibler (\KL) divergence loss function to \BCE\ helps capture more details of the pixel distributions of the maps. The \KL\ loss function,
\begin{equation}\label{eq:kl}
KL\left( q(\mathbf{x}) \parallel p(\mathbf{y}) \right) = \mathbb{E}_{ q(\mathbf{x})} \left[ \log \frac{ q(\mathbf{x})}{p(\mathbf{y})} \right]
\end{equation}
is defined as the expectation value, $\mathbb{E}_{ q(\mathbf{x})}$, of the logarithm of the ratio of the probability of the input $q(x)$ to that of the output $p(y)$. 
We also tested the Tweedie loss function \citep[TW,][]{zhou2022tweedie}, which has been shown to be more suitable for data with a skewed probability distribution. The TW loss function is defined in
\begin{equation}\label{eq:tw}
    TW\left(x, y, p\right) = -x \frac{{y}^{1 - p}}{1 - p} + \frac{{y}^{2 - p}}{2 - p}.
\end{equation}
It depends on a parameter $p$ that determines what the probability distribution should look like. We find that $p=0.5$ returns best values for training on the maps. 



\section{Results} \label{sec:results}

In this section, we present the reconstructed maps resulting from training the \bestAD\ structure (\AD\ with a \LSR\ of $50\times50$ pixels and a \BCE+\KL\ loss function) on \allmaps\ \Gmap s covering the whole parameter space of \set. In general, we find that the \AD\ model is able to reconstruct all of the input maps with visually acceptable quality (we will assess this in more detail in \autoref{sec:metrics}).


\autoref{fig:compare_results} shows the reconstruction of a single \Gmap\ with parameters $\kappa = 0.3, {\it \gamma} = 0.5, {\it s} = 0.5 $ using the \bestAD\ architecture. The first column shows the unconvolved maps, and the other columns correspond to the convolution with a source size of 0.1 and 0.5 \RE, respectively. \autoref{fig:compare_results_zoom} shows a $100 \times 100$ pixels zoomed in region of \autoref{fig:compare_results} only for a small segment of $100\times100$ pixels from the full-size map indicated in \autoref{fig:compare_results} by a red square.

\begin{figure*}[ht!]
    \centering
    \includegraphics[width=0.95\textwidth]{ 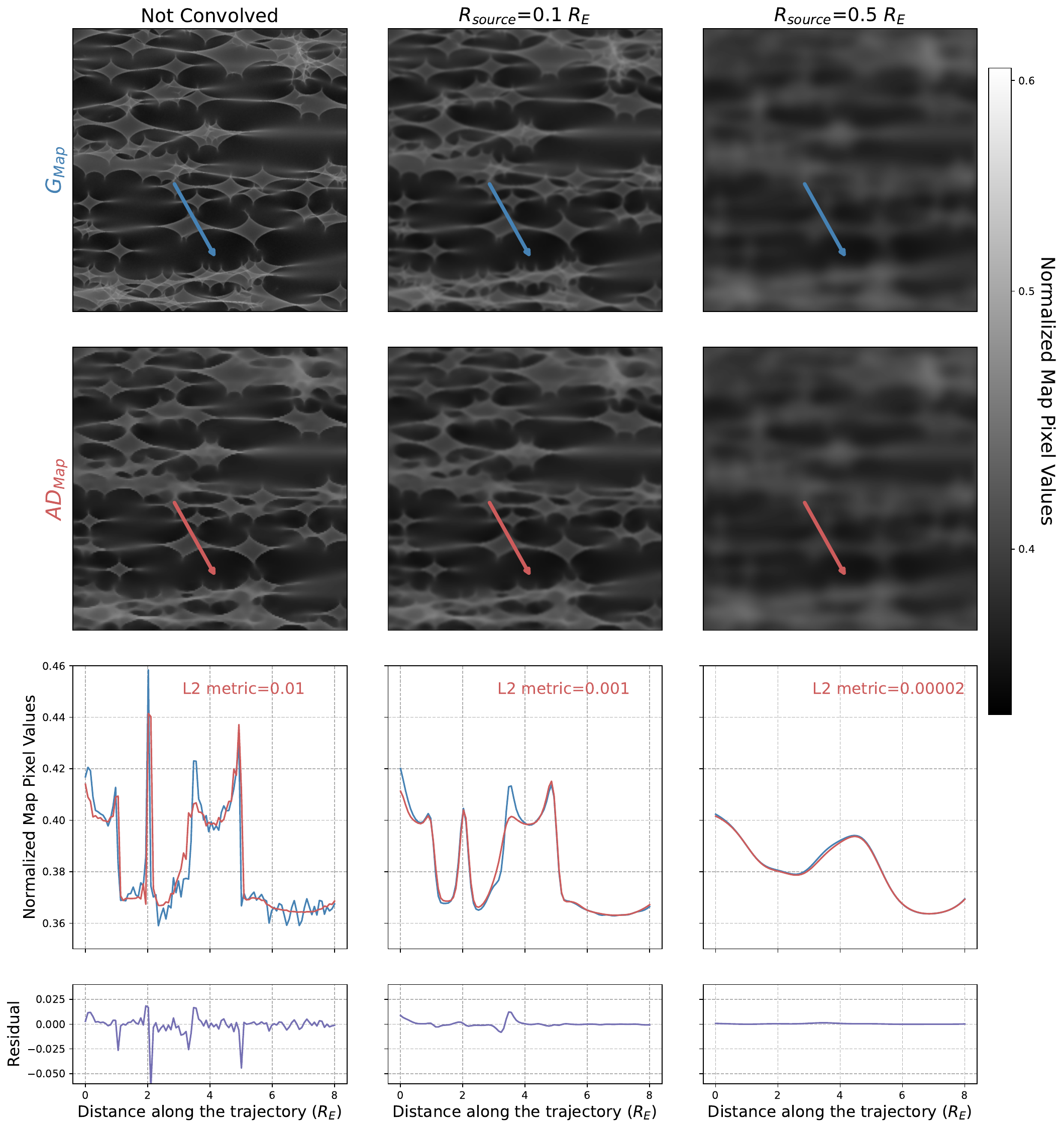}
    \caption{The first row shows the \Gmap s, and the second row shows the \ADmap s. The third row shows a light curve generated from the \Gmap\ (blue) and the \ADmap\ (red) along the same trajectory (red solid line on the maps), and the fourth row shows the residuals between the lightcurves. The $x$ axis of the third and fourth rows are shown in steps along the trajectory in units of Einstein radius \RE\ to avoid any assumptions for the velocity of the source along the trajectory. The $y$ axis of the light curves are in units of normalized map pixel values to show how the pixel values are reconstructed. Each column corresponds to a convolution status of the maps and their reconstructions so that the first column represents the unconvolved maps, and the other columns correspond to the convolution with a source size of $ R_{source} = $0.1, 0.5 \RE. The light curves from the original \Gmap\ (blue light curve in the third row first column of \autoref{fig:lc_examples}) exhibits some noise in its baseline. This noise comes from a Poisson-like noise in the original \ger\ maps that comes from counting rays per pixel \citep{vernardos2015gerlumph}. Note that the \AD\ has removed this noise from the \ADmap s. When the maps are convolved with a source of realistic size, the two light curves become very similar and their L2 metric printed on each panel decreases, and the blurring effects caused by the \AD\ process is hidden by the convolution. }
    \label{fig:lc_examples}
\end{figure*}

\subsection{Blurring of features in Autoencoder outputs}\label{sec:blurriness}
We see that the \AD\ has the effect of smoothing the sharper features in the un-convolved maps. This loss of resolution often happens with \AD s because they compress the input to a low dimension with a goal of preserving the most important information, and this results in the loss of high-frequency details and the blurriness of the outputs \citep{bank2020autoencoders}. In our case, in reality, the magnification maps are always convolved with a source that has a finite area. The effect of convolution with the source mitigates the smoothing effect of the \AD\, which becomes negligible.

\autoref{fig:lc_examples} compares an example of lightcurve generated from the same map as in \autoref{fig:compare_results} and its \AD\ version. In the third row, the light curves from the original \Gmap\ (blue light curve in the third row first column of \autoref{fig:lc_examples}) exhibits some noise in its baseline. This Poisson-like noise in the original \ger\ maps comes from counting rays per pixel \citep{vernardos2015gerlumph}. Note that the \AD\ has removed this noise from the \ADmap s. When the maps are convolved with a source of realistic size, the two light curves become very similar and their L2 metric (sum of square of differences between the two light curves) printed on each panel decreases (effectively indistinguishable, as we will also demonstrate in \autoref{sec:metrics}), and the blurring effect caused by the \AD\ process is hidden by the convolution. That is: under realistic observing conditions for the red end of the optical bands and larger wavelengths, where $ R_{source} \geq $ 0.1 \RE, the \ADmap s are an effective and accurate replacement for the \ger\ maps.

\subsection{Latent Space Representation}\label{sec:LSR}

The original images are generated by giving the parameters of the lens galaxy \set\ and the position of the compact objects. The \ADmap s can be generated using the \AD\ decoder given their \LSR. If the \AD\ is successfully trained, any two-dimensional array of size $50\times50$ with the correct range of values and patterns can lead to the generation of a new map when given to the decoder. It is therefore important to investigate the \LSR\ to understand how the information about \set\ parameters are preserved in this reduced dimension. 

\begin{figure*}[ht!]
    \centering
    \includegraphics[width=0.95\textwidth]{ 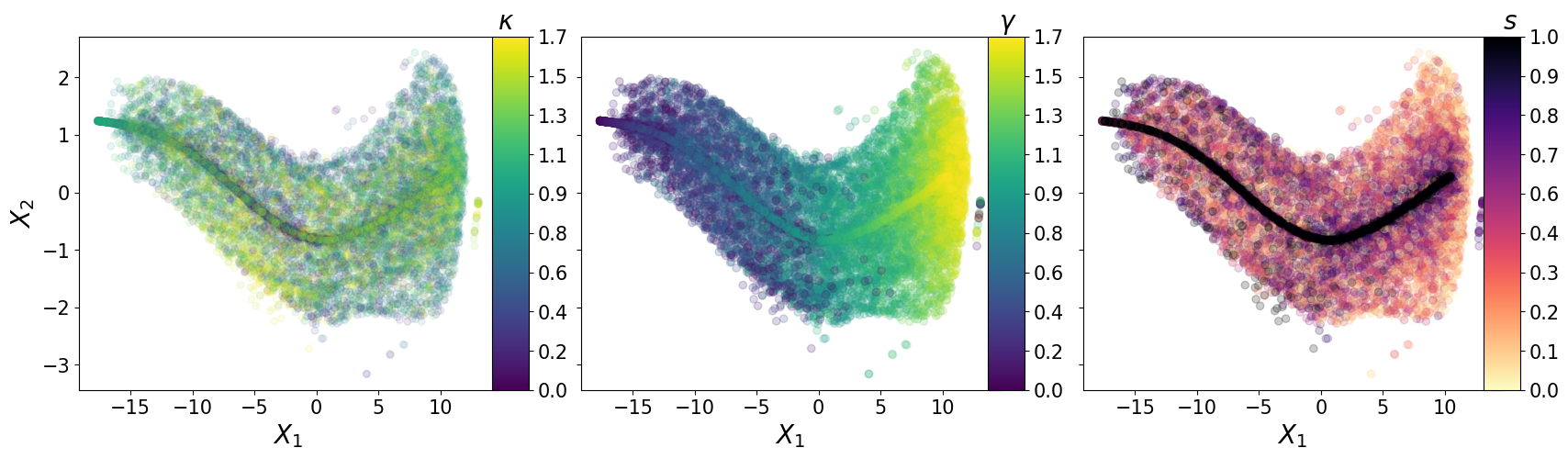}
    \includegraphics[width=0.95\textwidth]{ 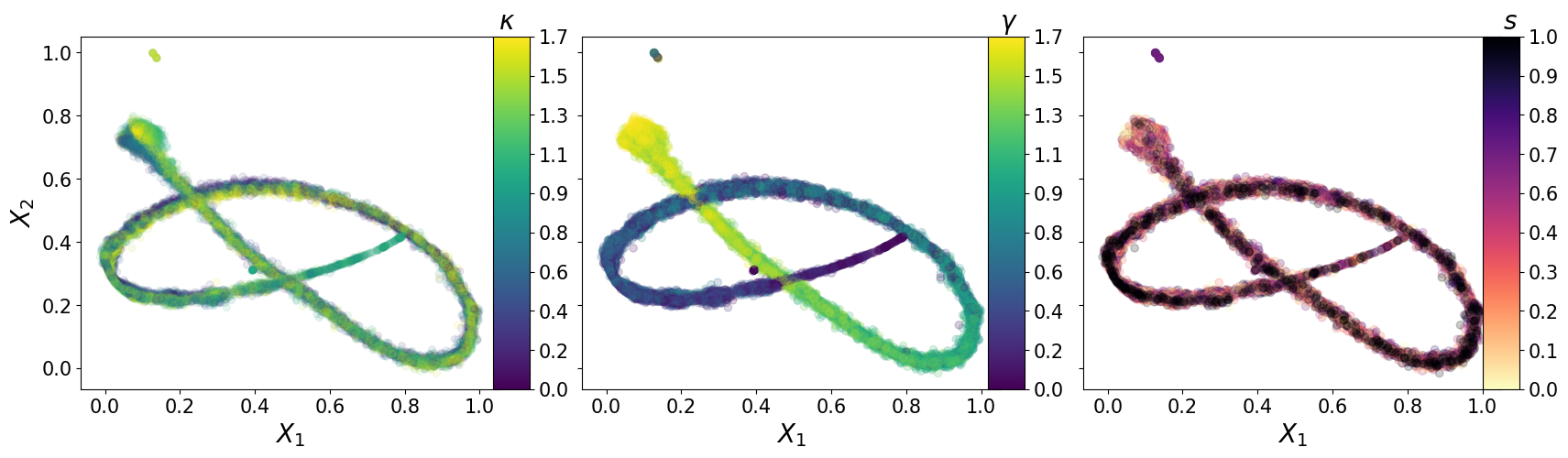}
        \includegraphics[width=0.95\textwidth]{ 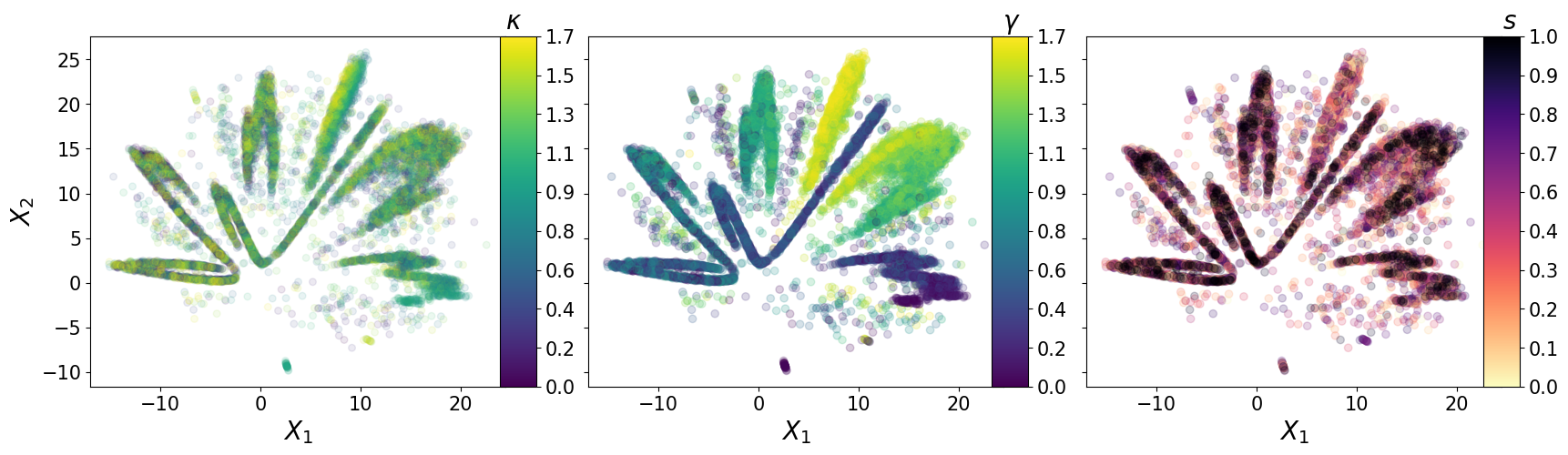}
    \caption{A two-dimensional visualization of the \LSR\ of all \allmaps\ \Gmap s in our training data are shown color-coded by \kp\ (left column), \g\ (middle column), and $s$ (right column). The first row is obtained by apply t-SNE dimensionality reduction approach to the \LSR. The second and third rows are obtained by applying UMAP to reduce the dimensions. Each row is independent of the others. Each point in each panel represents a single map in the sample. Patterns in color demonstrate that the information regarding the \set\ parameters are preserved in the \LSR\ that is 20 times lower in size compared to the original size. This figure is discussed in detail in \autoref{sec:LSR}. We also find that some of the outliers in the first row are also outliers in the second and third rows, and they correspond to maps that had some missing values.}
    \label{fig:LSR_tsnes}
\end{figure*}

\autoref{fig:LSR_tsnes} shows two-dimensional visualizations of the \LSR\ of our sample of \allmaps\ maps. Each point in the plots represents the \LSR\ of one map. Each column is color-coded by parameters \kp, \g, and $s$, respectively. The visualization in the first row is obtained by applying the t-distributed stochastic neighbor embedding algorithm \citep[t-SNE,][]{hinton2002stochastic, van2008visualizing} to the \LSR\ of the maps. The second and third rows are obtained using Uniform Manifold Approximation and Projection \citep[UMAP,][]{mcinnes2018umap} using a default Euclidean distance and a hyperboloid distance, respectively. UMAP reduces the dimension of the space with a focus on preserving local and global structure of the data whereas t-SNE has a focus on the local structure of the data. For t-SNE, the shape of the clustering in the low dimension is highly dependent on the hyperparameter choices (we chose $perplexity=500$ and $early\_exaggeration=2$) whereas UMAP is less sensitive to hyperparameter choices (for the second row, we chose the default distance $Euclidean$ and $n\_components=5$ where we used two of the components of the reduced dimension to plot in \autoref{fig:LSR_tsnes}; for the third row, we chose default $n\_components=2$ and $output\_metric=hyperboloid$. ).  We have selected these hyperparameters because they produced maps that were better segregated by \set. 

We see that the clustering in all three instances is driven by the values of \g, and a weaker correlation is seen with $s$ and \kp. In the third panel of the first row, we see a distinct black filament in the middle that corresponds to maps with $s=0.99$, meaning that for maps aligned along this filament, almost all of the mass of the lensing galaxy is in the form of smooth dark matter. \s=0.99 sequence is ordered entirely along the $x$-axis by \g. This separation, however, is not obvious in UMAP visualizations for the \s\ parameter. The correlation with  \kp\ is slightly more obvious in UMAP dimensionality reductions (first panel of the second row) as clusters or narrow regions with distinct values of \kp\ can be seen.

We investigated the nature of small clusters of maps separated from the main regions of the 2D distributions. For instance, we found that the set of maps separated into a cluster centered at $(x_1,x_2)\sim(12, 0.5)$ of the t-SNE visualizations are the same maps in the outlier blob of the second-row UMAP visualizations (top row of \autoref{fig:LSR_tsnes}) centered on $(x_1,x_2)\sim(0.2,1)$, and are outliers in the third-row as well (although distributed into a few separate clusters). We found that those were maps that had some artifacts (large number of missing pixel values). A cluster distinctly separated from the bulk of the maps in the third row (UMAP with hyperboloid distance), centered on  $(x_1, x_2)\sim (3,-10)$, has values \g $\sim$ 0, \kp $\sim$ 1. We note that these values are at the edge of the critical lines in the (\kp,\g) parameter space.

We argue that the \LSR\ has preserved information regarding the \set\ parameters. More detailed investigations of possible models are needed to connect the \LSR\ to the underlying parameters that we intend to pursue.


\section{Metrics}\label{sec:metrics}

Minimizing a loss function like the one in \bestAD\ throughout the training process results in reconstructing pixel values (BCE) and also reconstructing the probability distribution of the input map values (KL). Depending on the nature of the problem and the features of the input images, there might be other aspects of the reconstruction that are computationally expensive to include in the loss function. In our case, we need to demonstrate the \ADmap s can replace the original \ger\ maps. Therefore, we need to show that the \ADmap s and the light curves generated from them have the same statistical properties as the original \ger\ maps and their light curves, and eventually, they lead to identifying the same lensing properties given an observed lightcurve. 

We evaluate the quality of the reconstructed maps from three different aspects. First, we calculate the Fréchet Inception Distance \citep[FID,][]{frechet1906quelques} metric to evaluate the visual similarities of the \Gmap s and \ADmap s (\autoref{sec:FID_metric}). Next, we calculate an Anderson-Darling similarity test between the map value distributions of the \Gmap s and the \ADmap s called the Statistical Similarity Metric or SSM (\autoref{sec:sim_metric}). And finally, we assess the similarity of the light curves generated from the AD maps compared to the \Gmap s light curves which we discuss in \autoref{sec:lc_dist_metric}.

\subsection{FID metric}\label{sec:FID_metric}

The  Fréchet inception distance (FID) is commonly used to assess the quality and diversity of images generated by artificial intelligence generative models \citep{heusel2017gans}. In this approach, a pre-trained model (\FIDmodel, \citealt{szegedy2015going}) is applied to a real image set and to an image set generated by an Artificial Intelligence (AI) model, and a distance metric is used to evaluate the distribution of the features extracted from the real and AI-generated images, instead of comparing images one by one. The \FIDmodel\ model is a deep neural network trained on $\sim1$ million real-life images named ``ImageNet'' designed as a benchmark dataset for detection and classification \citep{russakovsky2015imagenet}. The features contained in the last layer of \FIDmodel\ have been shown to represent the quality and diversity of images. For example, when comparing a set of real images of human faces with a set of AI-generated images of human faces, the FID score is sensitive to the amount of noise and unreal features like dark spots in the fake images, along with distinguishing between images of human faces and any other images unlike a normal human face \citep{heusel2017gans}.

\begin{figure}
    \centering
    \includegraphics[width=0.5\textwidth]{ 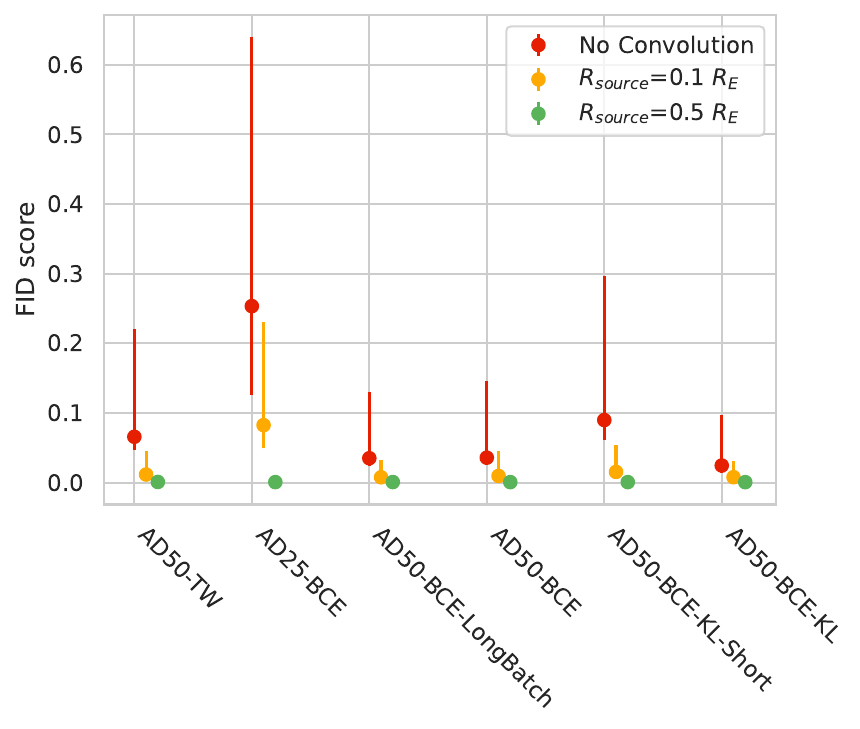}
    \caption{The FID metric, along with its confidence intervals, is shown for different \AD\ architecture (shown on the x axis). Different colors correspond to the convolution status of the maps, including when there is no convolution (red) and when convolving the \Gmap s and the \ADmap s with source radii of $ R_{source} = $0.1 (yellow) and 0.5 \RE\ (green). The metric is mostly similar among different models, and it becomes less sensitive to the model parameters when the maps are convolved. } 
    \label{fig:FID_metrics}
\end{figure}

To calculate the FID score for our  model, we apply the \FIDmodel\ to a set of 20 randomly chosen maps and their AD reconstructed versions. Please note that to use the \FIDmodel\ model, one needs to resize and preprocess the input images as required by the model. 
 Applying \FIDmodel\ to  $S_\mathrm{true}$  and $S_\mathrm{AD}$ ( $\displaystyle S_\mathrm{true}$ and $S_\mathrm{AD}$ are the  subsets of \Gmap\ and the corresponding \ADmap\ set), we have two sets of features $f(S_\mathrm{true})$ and $f(S_\mathrm{AD})$ with distribution $p_{true}$ and $p_{AD}$.
We then calculate the FID score (${d_f}^2$) using the mean and covariance matrices of the two distributions $p_{true}$ and $p_{AD}$ \begin{equation}\label{eq:fid_dist}
    \begin{split}
        {d_f}^2 = ||m_{true} – m_{AD}||^2 + \\
    Trace(C_{true} + C_{AD} – 2 \sqrt{C_{true} C_{AD}})
    \end{split}
\end{equation}
where $m_{true}$ and $m_{AD}$ are the means and $C_{true}$ and $C_{AD}$ are the covariance matrices of the two distributions respectively.
We repeat the calculation of the FID score 1,000 times each time for different sets of random\fd{ly selected} maps. We choose the median as the FID metric and calculate a confidence interval as the 16\% and 84\% percentiles of these values. The median FID metrics of six different AD architectures are shown in \autoref{fig:FID_metrics} along with their confidence interval. The model dubbed {\it AD25-BCE} (see \autoref{tab:trials}) underperforms other architectures, particularly \bestAD\, which is consistent with human interpretation based on the \ADmap\ and the \Gmap. We note that even though the FID metric can identify AD models that produce more faithful maps, we still cannot claim that the AD reconstructed maps do not contain any nonphysical features. We address these aspects with the other metrics discussed in the next two subsections.

\subsection{SSM: Statistical Similarity Metric}\label{sec:sim_metric}

While the FID metric is an established method for assessing the quality of machine-learning-generated replicas of true images, the properties of real-world images are rather different from the properties of caustic maps. For this reason, we further assess the output of out \AD\ with additional metrics designed to assess the statistical properties of the generated maps (this section) and, most importantly, whether they lead to the same scientific inference (\autoref{sec:lc_dist_metric}).

\citet{vernardos2013new} shows that the maps with the same \set\ have very similar distributions, visualized as histograms, over most of the parameter space, and these histograms are unique in small areas of the parameter space. In other words, the change in positions of the stars does not change the overall statistical properties of the maps. We therefore proceed to test the equivalence of the probability distributions of \Gmap s and \ADmap s with the following Statistical Similarity Metric (SSM). We choose 10,000 random maps and perform the Anderson-Darling test between the true map and the AD maps pixel value distributions. We do this with the original maps and when they are convolved with source radii of $0.1$ and $0.5 \; R_E$ (\autoref{sec:results}). We then take the median of the Anderson-Darling statistics as our similarity metric and measure a confidence range for it by taking the 16\% and 84\% percentiles. 

\begin{figure}
    \centering
    \includegraphics[width=0.5\textwidth]{ 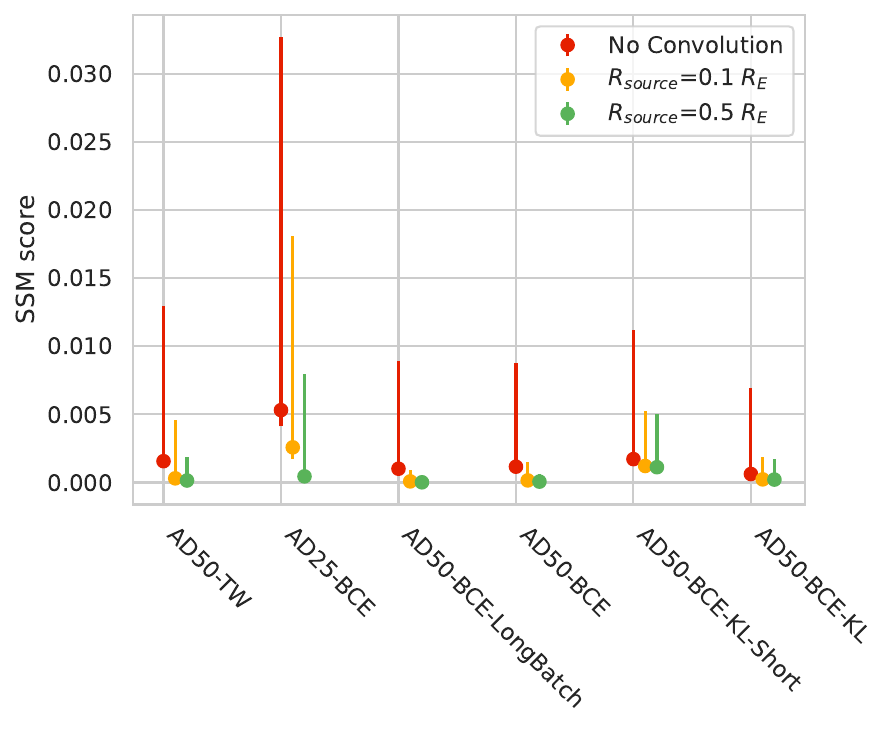}
    \caption{As \autoref{fig:FID_metrics}, but for the Statistical Similarity Metric (SSM).}
    \label{fig:sim_metrics}

\end{figure}

\autoref{fig:sim_metrics} shows the Statistical Similarity Metric for selected model architectures (see \autoref{tab:trials}) along with its confidence interval with different colors corresponding to original, un-convolved maps (red) and \Gmap s and the \ADmap s convolved with source radii of 0.1 (yellow) and 0.5 (green) \RE. This metric is pretty much independent of different \AD\ architectures and demonstrates that under expected observing conditions, any difference between the \Gmap s and \ADmap s is suppressed, effectively disappearing for an object with $R\geq0.5R_E$.

\subsection{Light Curve Distance Metric}\label{sec:lc_dist_metric}

While so far we demonstrated that there is a strong similarity between the \Gmap s and \ADmap s, to demonstrate their usefulness, we must demonstrate that in the context of scientific inference, replacing the \Gmap s with \ADmap s will lead to the same conclusions about the physical parameters of the system. In this section, we introduce the Light Curve Distance Metric. Note that this metric is not a one-to-one comparison between light curves from the \Gmap s and their \ADmap s. Previously, on \autoref{fig:lc_examples}, we showed a one-to-one comparison of the lightcurves with the same trajectory on the \Gmap\ and the \ADmap. The metric in this section is designed to evaluate the similarity between distributions of lightcurve minimum distances from comparing light curves with random trajectories on the \Gmap s and their \AD\ reconstructed version. 

We pick 1,000 \Gmap s with randomly chosen values of \set\ : $\mathrm{Gmap}_{m=1,...,1000}$. First, we measure the statistical similarity of light curves \emph{within} a given \Gmap.  For each \Gmap , $m$, we generate a light curve by choosing a random trajectory across the map (see \autoref{sec:intro}). This is our test-light curve \testLC$_{,m}$.  We then calculate the distance between \testLC$_{,m}$ and 1,000 light curves generated by choosing random trajectories across the same \Gmap\ ($GLC_{m,i=0,...,1000}$) as the L2 metric. We call $D_{GLC,m}$ the distance between \testLC$_{,m}$ and the closest (minimum L2) of the 1,000 $GLC_{m,i}$. 
\begin{equation}
    D_{GLC,m} (\testLC)=\\
    \min_i\{L2~(\mathrm{test}_{LC,m} - G_{LC,m,i})\}.
\end{equation}

For each of the 1,000 maps, we repeat this process 1,000 times, each time picking a new \testLC\ $l$ to compare to the same set of 1,000 random light curves from the same \Gmap , $m$. This results in a distribution $D_{GLCm}(l)$ of for each map, where the ${l}$ index stands for each \testLC. 

\begin{figure*}[ht!]
    \centering
    \includegraphics[width=0.9\textwidth]{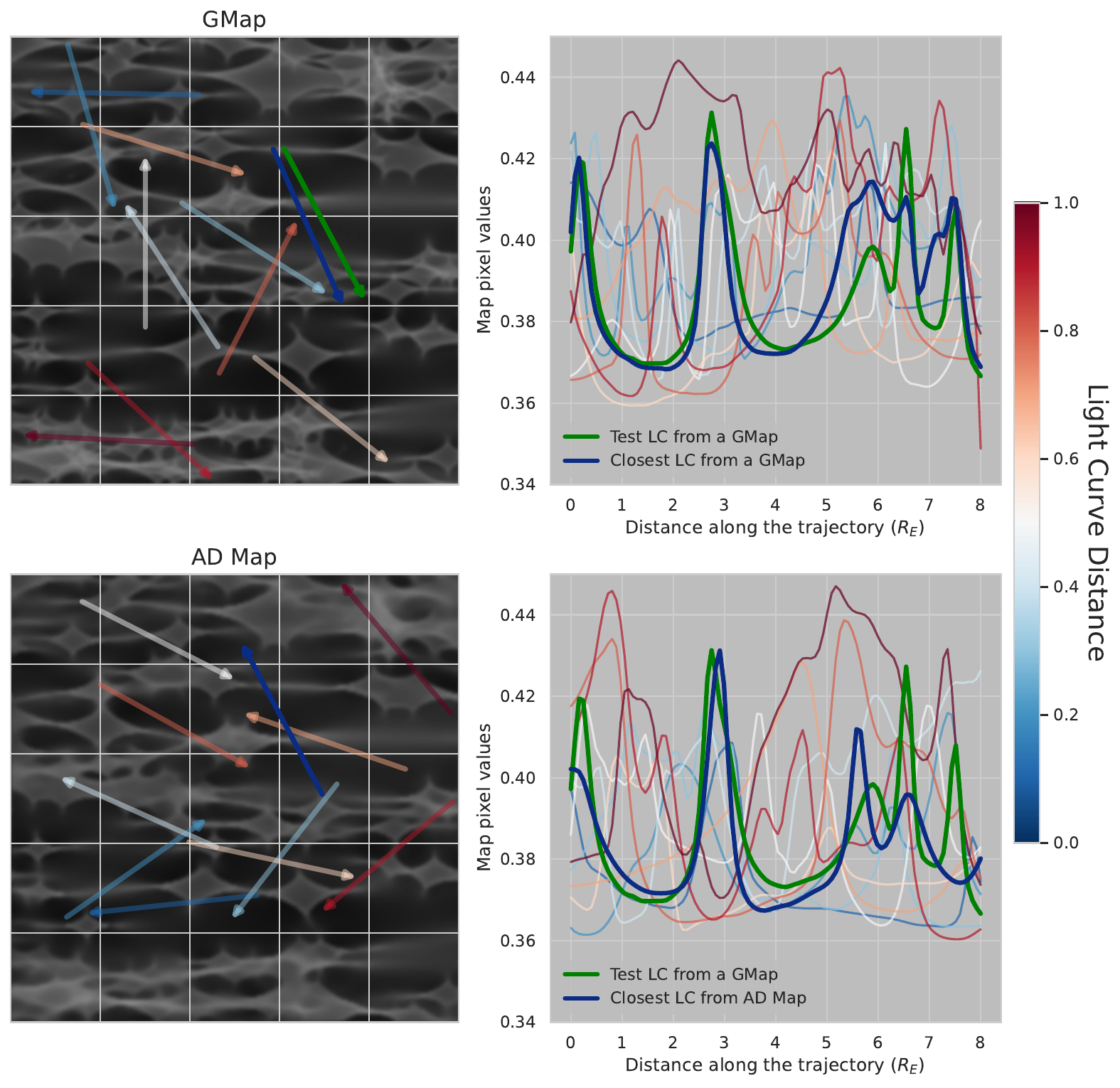}
    \caption{This figure shows examples of light curves extracted from \Gmap s and \ADmap s. The map chosen for this figure has a $\kappa = 0.3$, $\gamma=0.5$, and $s=0.5$. {\it Upper Left:} The \Gmap\ is plotted with trajectories of the generated light curve on top of it. {\it Upper Right:} The \testLC\ (see \autoref{sec:lc_dist_metric}) is shown in green, and its trajectory is also shown on the left panel in green. The light curve with minimum distance from the \testLC\ is shown in dark blue, along with its trajectory on the left panel. The other light curves are 10 examples of the 1,000 random light curves compared to \testLC. The colors represent the values of their L2 distance to \testLC\ chosen from each of 10 quantiles in the distribution of $D_{GLC,m}$ and $D_{ADLC,m}$. {\it Lower Left and Lower Right:} Same as the upper panels but for comparison light curves generated from trajectories across the \ADmap. Note that these light curves are generated by convolving a source of radius $ R_{source} = 0.1 \; R_E$.}
    \label{fig:min_lcs_example}

\end{figure*}

Next, we repeat the process but this time choose the comparison light curves from the \AD\ generated maps, rather than the \Gmap : $ADLC_{m,i}$. This leads to a distribution of minimum distances between the \testLC\ (extracted from the \Gmap) and comparison light curves extracted from the corresponding \ADmap s which we call $D_{ADLC,m}(l)$ for each \testLC\ $l$:

\begin{equation}
D_{ADLC,m}(\testLC)= \min_i\{L2~(\mathrm{test}_{LC,m} - AD_{LC,m,i})\}\end{equation},

\noindent
where the indices $m$ and $i$ have the same meaning as above. These two distributions can be compared to verify if the same physical parameters are inferred using the \Gmap s and the \ADmap s.
\autoref{fig:min_lcs_example} illustrates this process: we show a \testLC\ and a set of comparison trajectories and corresponding light curves. The top row shows the process applied to the \Gmap, and the bottom row shows the corresponding process applied to the \ADmap. 
The left panels show 
the map with the trajectories of the light curves plotted on top. The right panel shows the light curves. 
The \testLC\ and its trajectory are shown in green. The light curve most similar to \testLC\ is shown in dark blue along with its trajectory. Ten of the 1,000 random light curves compared to \testLC\ are shown color-coded by their distance to \testLC. These 10 lightcurves are chosen from each of the 10 quantiles in the distribution of $D_{GLC,m}$ and $D_{ADLC,m}$ at the top and bottom, respectively. 
Note that these light curves are generated by convolving a source of radius $0.1 \; R_E$. The comparison between random light curves in the true map and the AD map in \autoref{fig:min_lcs_example} indicates the same patterns of similarities.

\begin{figure*}[ht!]
    \centering
    \includegraphics[width=0.9\textwidth]{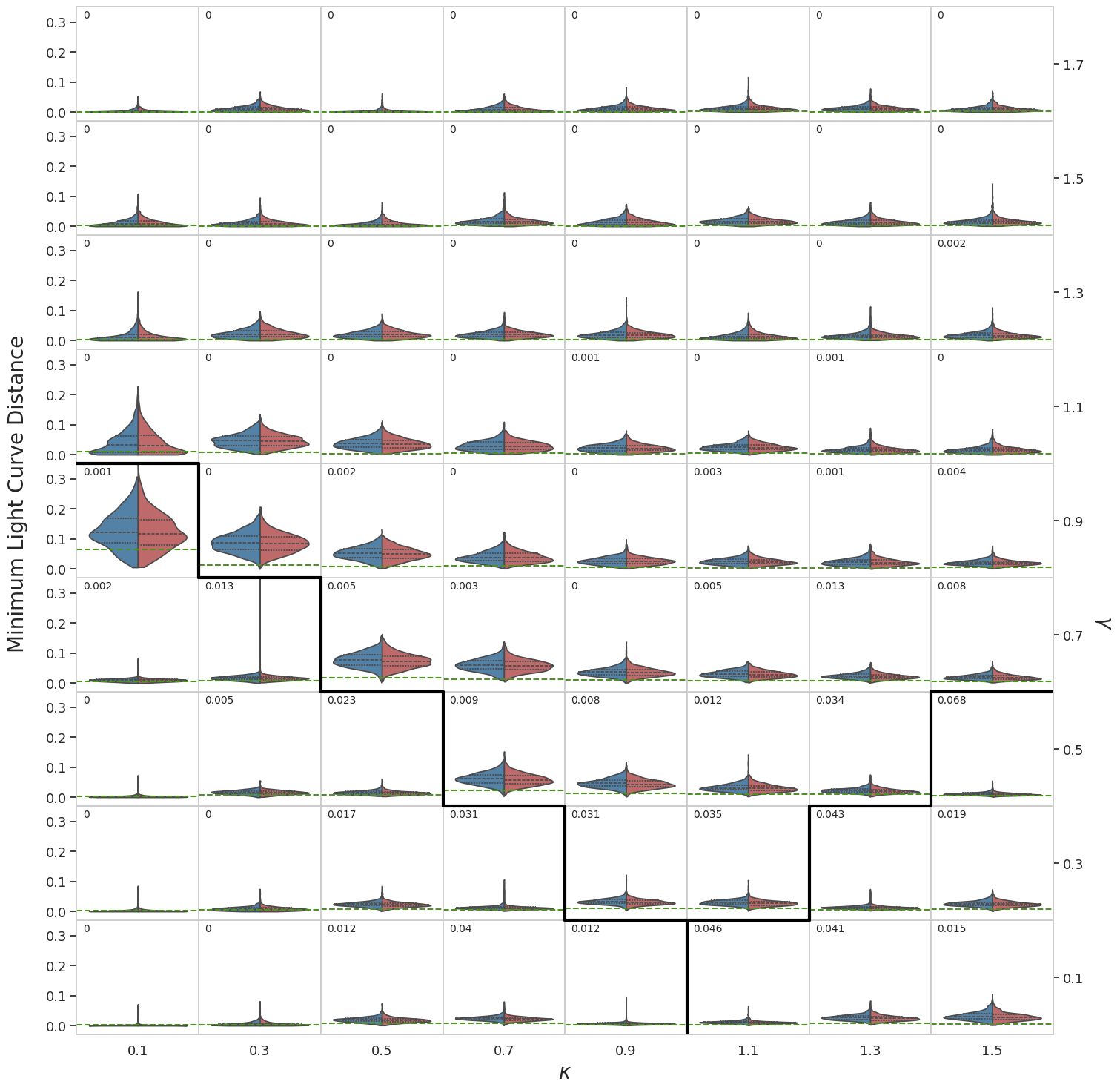}
    \caption{For 72 maps across the parameter space of \kp\ and \g\ and for $s=0.5$, we compare the distributions $D_{GLC,m}(l)$ (in blue) and $D_{ADLC,m}(l)$ (in red) of minimum light curves distances for the \Gmap s and the \ADmap s using a violin plot for each map. The median (black dashed line) and the 25\% and 75\% quartiles (black dotted lines) of the two distributions are also shown. The range of the y-axis is the same for all the violin plots and is shown on the left side of the plot. The values of \kp\ and \g\ are shown on the x and y axis, respectively. Even though the range of light curve distances is different for different maps, the distributions are almost identical between the \Gmap s and the \ADmap s. The green dashed line in each subplot represents the maximum light curve distance between one-to-one light curves of the \Gmap s and \ADmap s. The thick black line separating the panels indicates the critical line that divides the \kp -\g\ parameter space into three areas corresponding to the minimum (below and left), maximum (below and right), and saddle-point (above) images. The values printed on the inset plots are the  Anderson-Darling statistics calculated for each pair of distributions. The closer to zero the values are the more similar the pair of distributions are to each other. Note that the light curves generated for this analysis are created by convolving the maps with a source of radius $ R_{source} = 0.1 \; R_E$. We also repeated this plot for other values of $s$, and we found that the two distributions are again very similar for each map.}
    \label{fig:min_lcs_violin}
\end{figure*}

\autoref{fig:min_lcs_violin} shows a comparison between the two distributions $D_{GLC,m}$ and $D_{ADLC,m}$ for 72 different maps ($m$) across the $\kappa-\gamma$ space and for $s=0.5$. The values of \kp\ and $\gamma$ and $s$ are shown on top of the inset plots. This figure shows that the distributions of minimum light curve distances are very similar between the \Gmap s (in blue) and the \ADmap s (in red). The green dashed lines in each subplot are the maximum L2 distance of one-to-one comparison of light curves from the \Gmap\ and the \ADmap. As seen in the plot, these values are much lower than the values of the L2 metric between light curves of random trajectories which is consistent with our expectation.
The light curves generated for this analysis are created by convolving the maps with a source of radius $0.1 \; R_E$. We also generated the same plot for other values of \s\ and saw the same similarity between the two distributions for each map.

Since the \set\ values are inferred for an observed light curve by finding the closest light curve generated by drawing trajectory across caustic maps, this leads to the conclusion that the \ADmap s can be used in place of the original \Gmap\ with no impact on the scientific inference.

\section{Computational performance and map resolution}\label{sec:performance}

As discussed in \autoref{sec:intro}, one of the reasons why the \ger\ maps have had such a significant impact in the field is their richness and size, at $10,000\times10,000$ pixels at a physical scale of $25~ R_E$. However, so far, we have only demonstrated we can reproduce a low resolution $1,000\times1,000$ pixel version of the \ger\ maps (which we dubbed \Gmap s). Here, we demonstrate that our work extends naturally to reproducing the original resolution \ger\ maps, and we report explicitly on the computational performance of our model as compared to traditional methods. 

Generating the magnification maps by ray tracing has a high computational cost \citep{bate2011graphics, 2012ApJ...744...90B, vernardos2014adventures,zheng2022improved} which scales rapidly not only with the number of pixels, but also on the number of stars within the field. Recently, \citet{zheng2022improved} introduced a method optimized to run on GPUs that improves the time for the generation of magnification maps. When simulating $N_s=10^4$  stars, they report their GPU-based ray-shooting code takes $\sim1$ seconds for the generation of $1,000\times1,000$ pixel and $\sim10$ seconds for $10,000\times10,000$ pixel maps compared with $\sim50$ and $\sim500$ seconds respectively for the \ger\ maps  (see figure 7 in \citealt{zheng2022improved}). 
Importantly, the time increases near-linearly with the number of stars, so that the generation of $1,000\times1,000$ pixel maps with $N_s=10^6$ stars requires more than an order of magnitude more time for both codes (see figure 6 in \citealt{zheng2022improved}). 
Conversely, our trained \AD\ generates $1,000\times1,000$ pixel maps in about $t_{CPU}\sim0.3$ seconds on a CPU, $t_{GPU}\sim0.3$ seconds on a T4 GPU and $t_{GPU}\sim0.1$ seconds on a A100 GPU with negligible RAM usage and, importantly, regardless of the number of stars. This demonstrated that our methods are indeed performing better than existing methods at this image size in all cases.

To address the resolution, we tested our \AD\ trained on \Gmap\ maps on the prediction of full resolution $1,000\times1,000$ segments of the original $10,000\times10,000$ pixel \ger\ maps. We find that the reconstruction performance is consistent with the performance measured on the generation of \Gmap s, even though the model was not trained on the high resolution version, with FID and SSM producing similar values on the high resolution as those reported in \autoref{sec:FID_metric} and \autoref{sec:sim_metric} for the reproduction of \Gmap s.

To generate the full size, full resolution $10,000\times10,000$ pixel maps, we would need to expand the network to include more convolutional layers. While this would not require a much higher GPU RAM, it would require much higher system RAM for training, which is not available to our group at the moment. Once trained, such a model would be ready to use on a CPU or a GPU with minimal RAM requirement.

\section{Summary and Conclusion} \label{sec:concl}

The magnification maps are two-dimensional representations of how the light from an extragalactic source like a quasar would be modified if it goes through a small volume of the lens galaxy. These maps are generated by assuming a set of \set\ along with random star positions and a stellar mass function and exhibit a variety of features as in examples of \autoref{fig:map_examples}.

Extracting parameters of the lens galaxy and quasar's structure, along with accurate measurement of time delays between different images of a lensed quasar, requires understanding the microlensing variability in the light curves of images of a lensed quasar. Magnification maps are an essential part of this analysis, but they are computationally expensive to generate. Fast and efficient search through the parameter space of \set\ requires fast generation of the magnification maps. To facilitate this. We have successfully built an autoencoder, \AD, (\autoref{sec:AD}), a deep-learning model, and trained it on \allmaps\ images of \ger\ magnification maps \citep{vernardos2014gerlumph} of reduced resolution and size ($1,000\times1,000$ pixels). The trained \AD\ can reconstruct maps from all over the parameter space of \set\ and generate a compressed version of the maps that contain the most important information in the maps at a minimal computational cost regardless of the number of stars simulated (\autoref{sec:performance}). 

We develop metrics to study various aspects of the maps and show that the reconstruction is reliable and does not create non-physical artifacts in the output maps and the light curves generated from them. The FID metric (\autoref{sec:FID_metric}), the Statistical Similarity Metric (SSM, \autoref{sec:sim_metric}), and the light curve distance metric (\autoref{sec:lc_dist_metric}) evaluate various aspect of the maps including the quality of existence of various visual features, the similarity between the probability distribution of the pixel values of the input and output metric, and the statistical features of the light curves generated from the \Gmap s and \ADmap s. These metrics show that most of the \AD\ architectures tested (see \autoref{tab:trials}) are performing similarly well on reconstructing the images, but in particular, bigger \LSR\ size, adding a KL function to the loss function (\autoref{sec:loss} and \autoref{eq:kl}), and training for more epochs (\autoref{sec:training}) helps capturing more details of the maps.

We observe a mild loss of resolution with \ADmap s that smooths the higher contrast regions of the maps and also smooths out sharper features in the light curves when there is no convolution with the source PSF. This is expected from the \AD\ because of the nature of the loss function that does not focus on the extreme wings of the pixel value distribution (\autoref{sec:blurriness}). On the other hand, we find that this effect is smaller than the smoothing effect of convolving an original map with a source of at least $R_{source} = 0.1 R_E$ ( \autoref{fig:lc_examples}). We show that these maps with their current resolution can be a replacement for the original \Gmap s when the source sizes are larger than $0.1 R_E$. This would make the maps suitable to be used in analysis of light curves in the red end of the optical spectrum and larger wavelengths, and particularly suitable for studying BLR of the quasars.

The best selected \AD\ architecture (\bestAD, \autoref{sec:results}) creates a \LSR\ of the maps with a size of $50\times50$ pixels and reconstructs the maps given its \LSR. We showed in \autoref{fig:LSR_tsnes} that the 2D visualization of the latent space of the maps has preserved information regarding the \set\ parameters. This is more obvious with \g\ in the second column of the plot, but is also visible for \kp\ especially in the second row, and with \s\ in the first row. We argue that the lower dimensional representation of the maps embedded by the \AD\ into the latent space can be connected directly to the underlying parameters of the lens galaxy to enable generating new maps, as we will show in a forthcoming separate paper.

We have tested the performance of the same trained \AD\ in the generation of $1,000\times1,000$ pixels segments of the original \ger\ maps at full resolution ($2.5\times2.5R_e$) and demonstrated that the \AD\ trained on low resolution \Gmap\ can reproduce high resolution \ger\ maps at the same fidelity as the low-resolution versions (\autoref{sec:performance}). Due to limited access to resources, we have not trained a model that can produce the full size and resolution $10,0000\times10,000$ \ger\ maps, but our tests demonstrate that this is feasible with small variations of the architecture we developed in this work by expanding the network to include more blocks of double convolutional layers. This would require large system RAM for training. Once trained, maps can be generated on regular CPUs and GPUs.
Studying and quantifying the performance of magnification map generation from a low dimensional space using artificial intelligence is an essential step in the roadmap to develop neural network-based models that can replace traditional feed-forward simulation at much lower computational costs.

\begin{acknowledgments}
We used the following software packages: \texttt{pandas} \citep{mckinney2011pandas},  \texttt{numpy} \citep{harris2020array}, \texttt{matplotlib} \citep{hunter2007matplotlib}, 
\texttt{scipy} \citep{2020SciPy-NMeth}, and
\texttt{tensorflow} \citep{abadi2016tensorflow}.

This publication was made possible through the support of an LSSTC Catalyst Fellowship to S.K., funded through Grant 62192 from the John Templeton Foundation to LSST Corporation. The opinions expressed in this publication are those of the authors and do not necessarily reflect the views of LSSTC or the John Templeton Foundation.
This work was performed on the OzSTAR national facility at Swinburne University of Technology. The OzSTAR program receives funding in part from the Astronomy National Collaborative Research Infrastructure Strategy (NCRIS) allocation provided by the Australian Government, and from the Victorian Higher Education State Investment Fund (VHESIF) provided by the Victorian Government.
\end{acknowledgments}

\vspace{5mm}

\bibliography{references}{}
\bibliographystyle{aasjournal}

\end{document}